\documentstyle[twocolumn,prb,aps,epsfig,floats]{revtex}
\begin{document}
\twocolumn[
\hsize\textwidth\columnwidth\hsize\csname@twocolumnfalse\endcsname
\draft
\draft
\title{Carrier dynamics and infrared-active phonons in
    $c$-axis oriented RuSr$_2$GdCu$_2$O$_8$ film}
\author{A.P.~Litvinchuk$^{a,\ast}$, S.Y.~Chen$^a$, M.N.~Iliev$^a$,
        C.L. Chen$^a$, C.W. Chu$^{a,b}$, and V.N. Popov$^c$}
\address{$^a$Texas Center for Superconductivity and Department of Physics,
University of Houston, Houston, Texas~77204-5002\\
$^b$Lawrence Berkeley National Laboratory, 1 Cyclotron Road, Berkeley,
California~94720\\
$^c$Faculty of Physics, University of Sofia, BG~1126 Sofia, Bulgaria}
\maketitle
\begin{abstract}
The conductivity spectra of $c$-axis oriented thin RuSr$_2$GdCu$_2$O$_8$
film on SrTiO$_3$ substrate, prepared by pulsed-laser deposition,
are obtained from the analysis of the reflectivity spectra over
broad frequency range and temperatures between 10 and 300 K.
The free charge carriers are found to be strongly overdamped with
their scattering rate (1.0 eV at room temperature)
exceeding the plasma frequency (0.55 eV).
Four phonon lines are identified in the experimental spectra and
assigned to the specific oxygen related in-plane polarized
vibrations based on the comparison with the results of a lattice
dynamics shell model calculations.
\end{abstract}
]
  
Thin films of ruthenium oxides are of increasing interest because
of their unique physical properties. Conducting Sr$_{1-x}$Ca$_x$RuO$_3$
materials, for instance, are known to be an excellent component
for the manufacturing of multilayers, which consist of conductors,
superconductors, and insulators.\cite{cava,bozovic}
Their magnetic properties could be varied in a controllable way
by altering the Sr-Ca ratio. More recently another ruthenium oxide
material, RuSr$_2$GdCu$_2$O$_8$ (Ru1212), is in focus of both experimental
and theoretical studies as it exhibits highly unusual combination of
magnetic and superconducting properties (see, e.g.
Ref.\onlinecite{tallon,mcl,chu,ting,nak} for a review). In this paper we
report the results of an optical study of $c$-axis oriented Ru1212 film.

A KrF excimer pulsed laser with wavelength  $\lambda$ = 248~{\it nm}
was used to deposit RuSr$_2$GdCu$_2$O$_8$ thin films
on (001)-oriented SrTiO$_3$ (STO) substrates. Stoichiometric Ru1212
target was prepared using standard solid-state reaction similar
to those reported earlier for the bulk sample.\cite{preparation}
The films were deposited under various growth conditions.
We found that, in agreement with the results of McCrone et al.\cite{mcc},
the Ru1212 phase can only be formed after high temperature
post-annealing.  Magnetization measurements revealed the existence of the
magnetic ordering transition at T$_{Curie} \approx $120~K, slightly lower
compared to bulk materials. The resistivity data did not show clear signs of
superconductivity down to 10~K. The details of film preparation will
be published elsewhere.\cite{chen}

\begin{figure}
\centerline{\epsfig{file=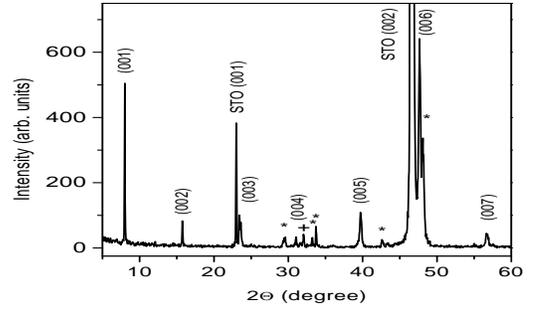,height=7cm,width=\linewidth,angle=0}}
\caption{X-ray diffraction scan for the annealed
RuSr$_2$GdCu$_2$O$_8$/SrTiO$_3$ film. The peaks
marked with an asterisk are due to impurity phase.}
\label{fig.1}
\end{figure}

Fig. 1 shows the typical X-ray diffraction scan pattern for the
annealed film. Pronounced ({\it 00l}) reflection peaks
indicate that the film is predominantly $c$-axis oriented.
This fact is also confirmed by our polarized Raman scattering studies
and comparison with the data obtained for Ru1212 single crystals.\cite{Raman}
The $c$-axis lattice parameter is determined to be 11.3~\AA,
in close agreement with earlier reports for sintered polycrystalline
samples.\cite{preparation,bauer}  The film is, though, not
phase pure as yet. The additional peaks (their intensity
did not exceed 10\% of the maximum peak intensity of Ru1212)
are marked with asterisks in Fig.~1 and originate from an unidentified
impurity phase. The peak marked with a "+" is due probably to the
(103)/(110) reflections of STO substrate.

Optical properties of Ru1212 film were studied
on a Bomem DA-8 Fourier Transform Interferometer in the frequency
range 50 - 10,000 cm$^{-1}$. A helium-flow cryostat was used for
temperature measurements between 300 and 10~K.
As the film thickness is rather small (0.3 $\mu$m),
one needs to account for the substrate response in
order to quantitatively evaluate the intrinsic properties of the film
from the reflectivity measurements. This could be done by taking into
account multiple reflections in the film/substrate system
(see, e.g., Ref.~\onlinecite{huml93}).

First, we present the results of model calculations of the reflectivity
for a conducting film on top of the SrTiO$_3$ substrate.
The dielectric function of either the substrate or the film as a
function of frequency $\omega$ was approximated as follows

\begin{equation}
\epsilon(\omega) = \epsilon_{\infty} - {S\over{\omega^2+i\omega\Gamma}}
    +\sum_i{F_i\omega^2_{0i}\over{\omega^2_{0i}-\omega^2-i\omega\gamma_i}},
\end{equation}

\noindent
where $\epsilon_{\infty}$ is the high frequency dielectric constant,
the Drude term accounts for the free carriers contribution with $\sqrt S$
and $\Gamma$ being the unscreened plasma frequency and scattering rate,
and the sum accounts for phonons with frequency $\omega_{0i}$,
damping $\gamma_i$ and  oscillator strength $F_i$.

The parameters of the three substrate oscillators
were obtained from the analysis of the reflectivity of the substrate
alone and their room temperature values are listed in Table I.
The high frequency dielectric constant is found to be
$\epsilon_{\infty} = 5.6$. All substrate parameters are
in close agreement with the reported data
(see, e.g., Ref.~\onlinecite{huml93}).

\begin{figure}
\centerline{\epsfig{file=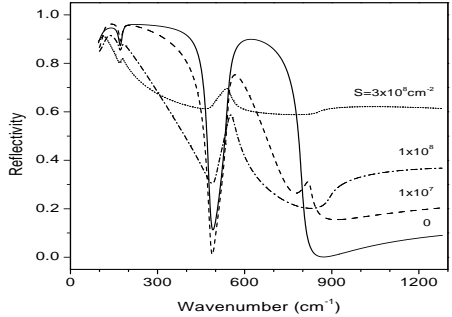,height=8cm,width=\linewidth,angle=0}}
\caption{Solid line is the reflection spectrum of bulk
SrTiO$_3$. Three other curves are simulated reflection
spectra of a 0.3~$\mu$m thick conducting film with $\epsilon_{\infty}=7.0$
on SrTiO$_3$ substrate.
The free carrier parameters of the film are $\Gamma$ = 7000 cm$^{-1}$
and is $S$ varying between $1 \times 10^7$ and $3 \times 10^8$ cm$^{-1}$
as indicated.}
\label{fig.2}
\end{figure}

\begin{table*}
\caption []{Phonon parameters of the SrTiO$_3$ substrate at room
temperature.}
\begin{tabular}{ c c c }
Frequency ($\omega_0$, cm$^{-1}$) & Damping ($\gamma$, cm$^{-1}$) &
Oscillator strength ($F$)  \\
\hline
109.2 & 19.3 & 209\\
175.4 & 13.4 & 5.52\\
542.1 & 20.3 & 1.46\\
\end{tabular}
\end{table*}
The experimental reflection spectrum of the substrate is shown 
by the solid line in Fig. 2. The other model curves in this Figure 
correspond to the substrate
covered by a 0.3 $\mu$m  thick film with $\epsilon_{\infty}$ = 7.0,
a fixed  plasma damping $\Gamma$ = 7000 cm$^{-1}$, and $S$ varying
from $1 \times 10^7$  to $3 \times 10^8$ cm$^{-2}$. The film with
the lowest $S$ weakly modifies the reflection
spectrum as the sharp minima at about 175 and 490 cm$^{-1}$
remain very pronounced; the reflectivity level increases, however, 
at  frequencies above 800 cm$^{-1}$.
An increase of the free carrier density in the film
($S = 1 \times 10^8$ cm$^{-2}$) affects considerably the shape
of the spectrum: the reflectivity drops rather quickly with frequency
in the range  150-450 cm$^{-1}$ and pronounced "restrahlen" band in the
range 500-800 cm$^{-1}$ becomes much narrower.
For $S = 3 \times 10^8$ cm$^{-2}$ the only substrate feature left in
the spectrum is the peak around 500 cm$^{-1}$ on top of a flat and rather
high reflection background.

\begin{figure}
\centerline{\epsfig{file=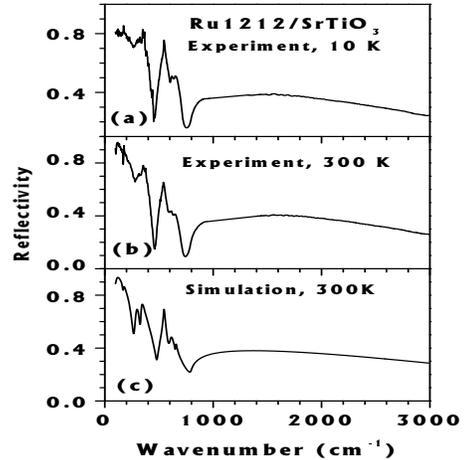,height=8cm,width=\linewidth,angle=0}}
\caption{Experimental reflection spectra of
RuSr$_2$GdCu$_2$O$_8$/SrTiO$_3$ film at 10~K (a) and 300~K (b). The lowest
panel (c) shows the simulated room temperature reflection of the
film/substrate system with the set of parameters listed in Tables I and II.}
\label{fig.3}
\end{figure}

Experimental spectra of film/substrate system at 10~K (the upper panel)
and 300~K (the middle panel) are shown in Fig.~3.
The fit of the room temperature spectrum (the lowest panel in Fig.~3)
reproduces all prominent features of the experimental spectra rather
well although we were not able to fit the spectrum within the noise level.
This fact implies that the Drude-Lorentz description of the free
carrier response in Ru1212 is oversimplified, but could be used
to get an estimate of free carrier plasma frequency and damping.
We mention that there are no drastic changes of the spectra upon cooling
down to 10~K.

The Ru1212 free carrier parameters at room temperature
obtained from the best fit are
$S$ = (1.2 $\pm$ 0.1) $\times 10^8$ cm$^{-2}$ and
$\Gamma = (8000 \pm 1000)$ cm$^{-1}$; the high frequency dielectric
constant $\epsilon_{\infty}$ is found to be 6.2 $\pm $ 0.2.
The screened plasma frequency of the film
$\omega_p = \sqrt {S/{\epsilon_{\infty}}}$ is
(4400 $\pm$ 1200) cm$^{-1}$, smaller than the scattering rate $\Gamma$,
a fact which points towards a strongly overdamped free carrier dynamics
in Ru1212.  One of the physical reasons of this high scattering rate might 
be the microstructure peculiarities, namely, the presence of well defined 
domains on the lateral scale 50-200 \AA,
as shown by the high resolution electron microscopy\cite{mcl}.
It appears that the dense domain structure is also at the origin of
complicated magnetic and superconducting properties of Ru1212.
Extremely high scattering rate and, correspondingly, very short carrier 
free path could also be due to structural or magnetic features of Ru1212 
on even a smaller scale, comparable to the lattice parameters.

\begin{table*}
\caption []{Parameters of the most intensive infrared active phonons
in Ru1212 at room temperature for polarization ${\vec E} \perp c$. 
The free carried parameters of the film are 
S = $1.2 \times 10^8$ cm$^{-1}$, $\Gamma$ = 8000 cm$^{-1}$.
$\epsilon_{\infty}$ = 6.2.}
\begin{tabular}{ c c c }
Frequency ($\omega_0$, cm$^{-1}$) & Damping ($\gamma$, cm$^{-1}$) &
Oscillator strength ($F$)  \\
\hline
285 &   32 & 10.2 \\
335 &   19 & 3.56 \\
604 &   40 & 0.74\\
655 &   10 & 0.07\\
\end{tabular}
\end{table*}
                                                              
Finally, relatively sharp structures in the reflection spectra of the Ru1212
film are due to the infrared active phonons with polarization
${\vec E} \perp c$. The fit of the spectra yields the set of phonon
parameters, listed in Table II.

\begin{figure}
\centerline{\epsfig{file=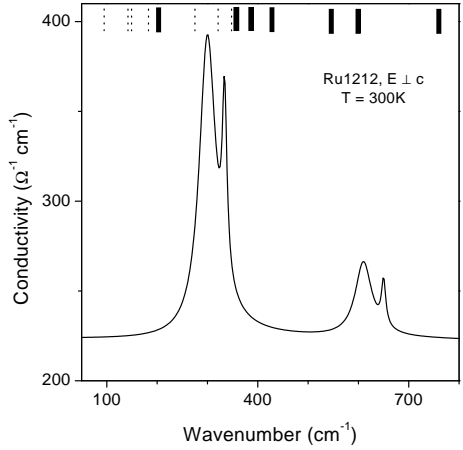,width=\linewidth,angle=0}}
\caption{Calculated spectral dependence of conductivity in
RuSr$_2$GdCu$_2$O$_8$ for ${\vec E} \perp c$ at room temperature.
The position of the transverse optical phonon modes,
as obtained from lattice dynamics calculation, is shown in the upper part
of the Figure by the vertical lines. Dotted lines correspond to the weak
phonons, while solid thick lines show the most intense modes.}
\label{fig.4}
\end{figure}

The conductivity spectrum of Ru1212, calculated according to the Eq.~1
and parameters from Table II is shown in Fig.~4. The fact of strongly
overdamped carrier dynamics is reflected in almost constant
in-plane "background" conductivity in the frequency range below
800 cm$^{-1}$. We mention that the absolute values of conductivity are 
by factor 
of two higher compared to what was earlier reported for ceramic
Ru1212 samples,\cite{ru1212cer,boris} for which the in-plane conductivity
was supposed to be a dominant component.\cite{gin}
The four peaks are the most intensive phonons.
In the upper part of Fig.~4 the position of the transverse optical (TO)
modes of Ru1212 for ${\vec E} \perp c$ is shown, calculated within
the shell model which, as we showed earlier, describes well $c$-axis
polarized modes in this material.\cite{ru1212cer}
For Ru1212, which crystallizes in the tetragonal $P4/mbm$ space
group\cite{chmais}, there are in total 14E$_u$ infrared-active modes
for light polarized normal to the $c$-axis.
Seven of them are predicted to possess vanishing oscillator strength
(LO-TO splitting is smaller than 3 cm$^{-1}$) and their position is
shown by thin dotted lines in the upper part of Fig.~4.
A number of more intense band are expected to appear between
350 and 400 cm$^{-1}$ (middle frequency range)
and 530-600 cm$^{-1}$ (high frequency range),
which are shown by solid lines in Fig. 4.
Experimentally, four phonon lines (parameters are listed in Table II)
are dominating the low frequency conductivity spectra of Ru1212.
By comparison with the results of lattice dynamics calculations,
they originate (in order of increasing frequency) from two apex
oxygen vibrations, oxygen in the Ru-O layer (O$_{Ru}$) and
Cu-O plane (O$_{Cu}$).

In conclusion, the frequency dependent conductivity of Ru1212 film
for ${\vec E} \perp c$ is obtained from the reflection data of thin
$c$-axis oriented Ru1212 film, deposited on SrTiO$_3$ substrate.
Free charge carriers of the film are found to be strongly overdamped.
The well developed domain structure of Ru1212 is a possible reason
for this effect. Four phonon lines have been identified in the spectra.
Based on the results of shell model lattice dynamics calculation, they are
assigned to the specific vibrations of oxygen in different crystallographic
positions.

{\bf Acknowledgements.} This work was supported by
NSF Grant No. DMR-9804325, the State of Texas trough the 
Texas Center for Superconductivity at the University of Houston,
T.L.L. Temple Foundation and John J. and Rebecca Moores Endowment.

\end{document}